\documentstyle[12pt]{article}

\input tcilatex
\QQQ{Language}{
American English
}

\begin{document}

\title{Some theoretical considerations concerning ion hydration in the case of ion
transfer between water and 1,2- dichloroethane.}
\author{C. S\'{a}nchez and E.Leiva\thanks{%
Corresponding author, e.mail: eleiva@fcq.uncor.edu} \\
Unidad de Matem\'{a}tica y F\'{\i}sica, Facultad de Ciencias Qu\'{\i}micas\\
Universidad Nacional de C\'{o}rdoba. Agencia Postal 4, C.C. 61\\
5000 C\'{o}rdoba,\ Argentina \and S.A. Dassie and A.M. Baruzzi \\
Depto. de F\'{\i}sico Qu\'{\i}mica, Facultad de Ciencias Qu\'{\i}micas\\
Universidad Nacional de C\'{o}rdoba. Agencia Postal 4, C.C. 61 \\
5000 C\'{o}rdoba, Argentina}
\maketitle

\begin{abstract}
Some aspects of direct ion transfer across the water/1,2- dichloroethane are
analyzed using a very simple model based on thermodynamic considerations. It
was concluded that ion solvation by water molecules may occur {\it in some
particular cases} in the organic phase, delivering an important contribution
to the Gibbs free energy of ion transfer between the aqueous and the organic
phase. In general terms, this particular type of transfer should be favored
in the case of highly charged small ions at interfaces with a relatively low
surface tension and a large difference between the reciprocal of the
corresponding dielectric constants.
\end{abstract}

\section{Introduction}

The understanding of charge transfer processes across liquid / liquid
interfaces is particularly important in many areas of chemistry, physical
chemistry and biology. In the last decades a variety of electrochemical
methods have been adapted to study these processes. Cyclic voltammetry,
chronopotentiometry, impedance spectroscopy applied to macro and more
recently to microinterfaces between two immiscible liquids are currently
being used \cite{MSK88}-\cite{Girault2}. Nevertheless, interpretation of the
experimental data requires in many cases a proper knowledge of the interface 
\cite{Kazarinov87}\cite{SMH84}. In recent years new experimental techniques
have been developed, which allow to obtain molecular level information
concerning the interface. Second harmonic generation at the liquid / liquid
interface provides information about the state of molecules, especially the
surfactants adsorbed at the interface\cite{CR95}\cite{CFZ96}\cite{NHC95}.
Scanning electrochemical microscopy studies of charge transfer processes
across ITIES are in progress\cite{RMB93}\cite{SB95}. The concentration
profiles on both sides of the phase boundary can be probed and the thickness
of the mixed solvent layer can be evaluated by scanning the
ultramicroelectrode tip\cite{WMB95}. More recently, IR reflection
spectroscopy\cite{SF90} and specular reflection of neutrons \cite{LLF91}
have been applied yielding information at the molecular level. Theoretical
calculations should provide an important contribution to give insight into
interfacial structure and the mechanism of transfer across ITIES; several
authors have used molecular dynamics to model the interface, the ion transfer%
\cite{Benjamin93}\cite{Benjamin94} and the water/oil interface in the
presence of micelles\cite{SHEROS91} or in the presence of a monolayer of
amphipatic molecules\cite{KKNNS92}. Benjamin \cite{Benjamin92} has employed
equilibrium free energy calculations and equilibrium dynamics trajectories
to provide a detailed microscopic picture of the ion transfer processes. He
showed that the transfer into the organic phase is accompanied by a
hydration shell of water molecules and may also involve ion-pairing. A
lattice gas model has been recently used by Pereira et al. \cite{Pereira}and
Schmickler \cite{Schmickler} to clarify a number of experimental features of
liquid-liquid interfaces, including capacity and ion transfer. On the other
hand, continuum models have also been employed to understand ion transfer
between immiscible liquids. A review on this topic has been given by Markin
and Volkov\cite{MV89}, where the Born model, non-linear dielectric effects
and the use of non-local electrostatic methods were analyzed. Although
quantitatively poor, the Born theory was found to provide a qualitative
understanding of ion transfer processes from an aqueous to an organic phase,
concerning the effect of static permittivity and ion size. Among continuum
models, it is also worth mentioning the contribution of Abraham et al. in a
sequence of papers\cite{A&L1}\cite{A&L2}\cite{A&L3}\cite{A&L4}\cite{A&L5},
where calculations of free energies and entropies of solvation were
performed for a number of systems. We shall refer to these papers as $A\&Ln$%
, the number $n$ denoting the number with which each part was originally
labelled.

It is the purpose of the present work to revive and shear new light on one
of the aspects considered in $A\&L5$, which is that of ion hydration in the
organic phase. We add a new feature to their model that introduces an extra
term to the free energy of transfer which was previously ignored . This
accounts for the interaction between the water layer surrounding the ion in
the organic phase and the organic solvent. We also use the Born model to
discuss some qualitative features of the present formulation.

\section{The model}

We consider the direct transfer of alkaline and alkaline earth ions,
indicating this process as:

\begin{equation}
M^{+z}(aq)\rightleftharpoons M^{+z}(org)  \label{n(x)}
\end{equation}

As pointed out in $A\&L5$\cite{A&L5} in their calculation of partition
coefficients, the free energy of transfer $\Delta G^t$ of the ion from water
to the organic solvent contains in principle neutral and electrostatic
terms. Since the former contributes to $\Delta G^t$ only to a minor extent,
we shall ignore it in most of the discussion given below.

Table I shows the Standard Gibbs energy $\Delta G_{tr}^0$ of this reaction
as given by Sabela et al. \cite{Sabela}, compared with the corresponding
values for several anions. In contrast to the transfer of anions, where
large differences are observed, the transfer energy for the smaller alkaline
ions appears to be rather size-insensitive. An interpretation of these
results could be attempted in terms of a continuum model of the solvent with
inclusion of non linear dielectric effects or non local electrostatic
models, as proposed in an interesting review by Markin and Volkov\cite{MV89}%
. However, we first want to draw attention here to an alternative model,
which is also based on a continuum model for the solvent, but introduces a
feature that has been absent in most of the previous theoretical
considerations. This characteristic is the allowance for a remaining
solvation sheet belonging to the aqueous phase {\it even after} the ion has
been transferred to the organic phase. In fact, although Abraham and Liszi 
\cite{A&L5} did consider the existence of a hydration sheet in the organic
phase, a relevant contribution to the free energy of transfer was omitted
there, as discussed below. At the very preliminary stage of development of
these speculations, we will not seek a close agreement with the experiment,
but rather point out under which circumstances this kind of events may
occur. More sophisticated theoretical considerations and further
experimental evidence will allow the confirmation or rebuttal of the present
ideas.

We turn now to consider a naive but enlightening model for the transfer of
ions from the aqueous to the organic phase and we later considered some
improvements to it within the continuum approach. In order to avoid
confusion, we want to state here what we mean when we write below the
''radius of the ion''. In the case where we use ideas related to the work of
Abraham and Liszi, we shall be talking about the crystallographic radius. On
the other hand, when we apply the concept developed by Rashin and Honig \cite
{Rashin&Honig85}, we shall be actually referring to the cavity that the ion
produces into the solvent. A detailed discussion about this point has been
given by Rashin and Honig\cite{Rashin&Honig85}. Our first simplified model
is shown in Figure 1 and follows the ideas suggested in $A\&L5$. We assume
that the alkaline ion of radius $r_o$ eventually {\it may} carry with it a
solvation sheet of radius $r_s-r_o$, so that after the transfer it is
immersed in a mixed dielectric medium, constituted by a sphere of dielectric
constant $\varepsilon _1$, the same as that of the aqueous phase, which is
in turn embedded in a dielectric continuum of constant $\varepsilon _2$. We
shall refer to this kind of ionic transfer as a transfer through a ''water
peel (WP)'' mechanism. If we calculate the electrostatic free energy change
for this process according to the spirit of Born's model, we get:

\begin{equation}
\Delta G_{elec}^{0\prime }=-\frac{q^2}{2r_0\varepsilon _1}+\frac{q^2}%
2\left\{ \frac 1{\varepsilon _1}\left[ \frac 1{r_0}-\frac 1{r_s}\right]
+\frac 1{\varepsilon _2r_s}\right\}  \label{delgmixp0}
\end{equation}
where $q$ represents charge of the ion. However, a very important element is
missing in this model, as can be understood from the following argument. The
presence of the aqueous solvation sheet makes the ion more stable in the
organic phase, and this stability increases as $r_s$ increases. Thus, on
electrostatic grounds $r_s$ should increase indefinitely to stabilize the
ion as much as possible. The reason for this not to happen is that the
creation of the water layer in the organic phase must have some cost, this
depending on the radius $r_s.$ Note that these considerations are absent in $%
A\&L5$, where the water layer is somewhat arbitrarily (but very reasonably)
assumed to be of one water molecular diameter. According to the previous
argument, we shall write the free energy of transfer according to:

\begin{equation}
\Delta G_{transf}^{0\prime }=-\frac{q^2}{2r_0\varepsilon _1}+\frac{q^2}%
2\left\{ \frac 1{\varepsilon _1}\left[ \frac 1{r_0}-\frac 1{r_s}\right]
+\frac 1{\varepsilon _2r_s}\right\} +4\pi \gamma r_s^2  \label{delgmixp}
\end{equation}
where $\gamma $ denotes the surface tension of a water/organic solvent
interface. Thus, the three terms on the r.h.s. of eqn. \ref{delgmixp}
represent the solvation energy of the ion in the aqueous phase, the
solvation energy in the mixed phase and the free energy required to create
the new surface originated by the water sphere of radius $r_s,$
respectively. The last term looks like the solvophobic term discussed by
Markin and Volkov \cite{MV89} when considering the contribution of the
solvophobic effect to the resolvation energy. However, it is clear that it
has here quite a different meaning, since $r_s$ is no longer the ionic
radius but that of the hydrated ion.

On the other hand, the transfer of an alkaline ion from the aqueous to the
organic phase without the aqueous solvation sheet would involve the
contribution:

\begin{equation}
\Delta G_{trans}^0=\frac{q^2}2\frac 1{r_0}\left[ \frac 1{\varepsilon
_2}-\frac 1{\varepsilon _1}\right]  \label{delgmix}
\end{equation}

We are aware that we actually should use instead of this equation a similar
one taking into account that the cavity radius produced by a particular ion
will vary in different solvents \cite{Rashin&Honig85}, but we postpone this
discussion for below.

Thus, the fact whether the ion will be transferred with an aqueous solvation
sheet or not, will depend on the relative magnitudes of the processes
involved in \ref{delgmixp} and \ref{delgmix}. Simplification of eqn. \ref
{delgmixp} yields:

\begin{equation}
\Delta G_{trans}^{0\prime }=\frac{q^2}2\frac 1{r_s}\left[ \frac
1{\varepsilon _2}-\frac 1{\varepsilon _1}\right] +4\pi \gamma r_s^2
\label{delgmixp2}
\end{equation}

Note that according to this expression $\Delta G_{trans}^{\prime }$ is
independent of the ionic radius, in agreement with the experimental
observation for the smaller alkaline ions. We will now analyze under which
conditions the processes related to \ref{delgmix} or \ref{delgmixp2} should
prevail. We illustrate this point in Figure 2, where we make a qualitative
plot of $\Delta G_{trans}$ and $\Delta G_{trans}^{\prime }$ as a function of
the radius of the aqueous sphere surrounding the ion. Three different
situations may appear. In the first (Figure 2a) the curves for $\Delta
G_{trans}^{\prime }$and $\Delta G_{trans}$ intersect in two points, and
there exists an infinity of values of $r_s$ which allow the inequality $%
\Delta G_{trans}^{\prime }<\Delta G_{trans}$. Alkaline ion WP transfer
should be the predominant mechanism for the ion transfer in this case. In
the second (Figure 2b), there is only one point in which we get $\Delta
G_{trans}^{\prime }=\Delta G_{trans}.$ In the third case, we get $\Delta
G_{trans}^{\prime }>\Delta G_{trans}$ for all $r_s$, so that ion WP transfer
should not occur.

The relevance of the physical parameters involved in the present model may
be envisaged from the following analysis. Let us set the necessary condition
for a minimum in eqn. \ref{delgmixp} according to:

\begin{equation}
\frac{d\Delta G_{trans}^{0\prime }}{dr_s}=-\frac{q^2}2\frac 1{r_s^2}\left[
\frac 1{\varepsilon _2}-\frac 1{\varepsilon _1}\right] +8\pi \gamma r_s=0
\label{deriv}
\end{equation}
which leads us to the value for the radius of the aqueous sphere at the
minimum:

\begin{equation}
r_s^{\min }=\frac 12\left[ \frac{q^2}{2\gamma \pi }\left( \frac
1{\varepsilon _2}-\frac 1{\varepsilon _1}\right) \right] ^{(1/3)}
\label{rsmin}
\end{equation}

Substituting $r_s^{\min }$ back into eqn. \ref{delgmixp2}, we get:

\begin{equation}
\Delta G_{trans}^{0\prime \text{ }\min }=\frac 32\frac{q^2}2\frac
1{r_s^{\min }}\left[ \frac 1{\varepsilon _2}-\frac 1{\varepsilon _1}\right]
\label{delgmin}
\end{equation}
which resembles in shape eqn. \ref{delgmix}. Thus, the existence of a
minimum for $\Delta G_{trans}^{\prime }$ located below $\Delta G_{trans}$
(the case depicted shown in Figure 2a) will imply the condition :

\begin{equation}
r_0<\frac 23r_s^{\min }  \label{r0rs}
\end{equation}

If we now substitute $r_s^{\min }$ into eqn. \ref{r0rs} and rearrange, we
arrive at the inequality:

\[
\frac{\gamma r_0^3}{q^2}<\frac 1{54\pi }\left[ \frac 1{\varepsilon _2}-\frac
1{\varepsilon _1}\right] 
\]

This is a very simple mathematical formulation emphasizing the relevance of
the physical magnitudes which will allow a WP transfer of ions from an
aqueous phase to an organic one: the surface tension of the corresponding
water/organic interface, the ionic radius and the charge of the ion. Thus,
for a given water/organic interface, the ionic transfer through the WP
mechanism will be favored for interfaces with the lower surface tension and
ions with the smaller radii and the higher charges. In order to make a
quantitative test of these ideas, we have plotted in Figure 3 the standard
Gibbs energies for the transfer of some univalent ions from water to
1,2-dichloroethane, obtained from electrochemical measurements\cite{Sabela}%
\cite{Samec88} as a function of the reciprocal of the ionic radius. In
Figure 3a we used the ionic radius as usually reported in textbooks\cite
{Bockris}, and in Figure 3b we employed the corrected radius as quoted by
Rashin and Honig \cite{Rashin&Honig85}. This corresponds to the choice of
ionic radii for anions and covalent radii for cations, all increased by 7
\%. Although there remain some differences between anions and cations, the
latter choice seems to be more adequate for this representation . In the
case of all the anions and the bigger cations, an increase of $\Delta
G_{trans}$ with $1/r_0$ is found, a fact which was predicted by eqn. \ref
{delgmix}. The differences observed between cations and anions may be
attributed to the fact of using the same $r_0$ irrespective of the solvent.
It has been pointed out in the literature that different radii for the
cavity surrounding the ions are expected, especially when the solvents
involved have a different polarity \cite{Rashin&Honig85}. On the other hand,
a remarkable deviation from the linear behavior is found in the case of the
smaller cations $Li^{+}$ and $Na^{+}$. Furthermore, the experimental $\Delta
G_{trans}$ is not found to change with the ionic radius, a feature which was
anticipated in the discussion given above by eqn. \ref{delgmin}. Due to the
simplicity of the present model, only a qualitative description of the
direct transfer can be pursued here. However, we can attempt an improvement
in terms of dielectric saturation effects as proposed in $A\&L3.$ Thus, eq. 
\ref{delgmixp2} can be replaced by:

\begin{equation}
\Delta G_{trans}^{0\prime }=\frac{q^2}2\dint\limits_{r_0}^\infty \frac
1{r^2}\left[ \frac 1{\varepsilon _2(r)}-\frac 1{\varepsilon _1(r)}\right]
dr+4\pi \gamma r_s^2  \label{delgmixp3}
\end{equation}
where $\varepsilon _1(r)$ and $\varepsilon _2(r)$ are now electric field
dependent functions as given in $A\&L3$. These can be calculated numerically
by solving the equation $\varepsilon =n^2+\frac{4\pi \rho \mu }E\left( \coth
(\frac{3\mu E}{2kT})-\frac{2kT}{3\mu E}\right) $ along with $E=\frac
q{\varepsilon \ r^2}$ , where $n$ denotes internal refractivity of the
liquid, $\rho $ the number density, $\mu $ the dipole moment and $kT$ has
the usual meaning. The integral can also be performed numerically and
numerical minimization of \ref{delgmixp3} provides an estimation of $%
r_s^{\min }$ and $\Delta G_{trans}^{\prime \text{ }\min }$. These are given
in Table 2 along with the more approximate values stemming from Born's
theory. These values point towards a hydration sheet close to a molecular
diameter, as suggested in $A\&L5$ and the value for $\Delta
G_{trans}^{\prime }$ reasonably agrees with the experimental finding for $%
Li^{+}$and $Na^{+}$, where we expect WP\ transfer to take place. Since there
is theoretical evidence that the surface tension of small drops should
decrease with increasing curvature\cite{Allen}, we expect from eq. \ref
{delgmixp3} that the hydration extent in the organic phase should be
somewhat larger than what we calculate here. However, the minimum described
by this equation is relatively flat so that important changes in the
hydration extent should not appreciably change the values of $\Delta
G_{trans}^{\prime }$we report here. A further comment is also relevant
concerning the accuracy of eq. \ref{delgmixp3}. We expect the values
obtained from it for WP transfer to be much more accurate that any
estimation of transfer free energy that can be made for ion transfer {\it %
without hydration. }In fact, if we assume that the structure of the
hydration sheet remains essentially unaltered in the organic phase, this
equation reduces to:

\begin{equation}
\Delta G_{trans}^{0\prime }=\frac{q^2}2\dint\limits_{r_s}^\infty \frac
1{r^2}\left[ \frac 1{\varepsilon _2(r)}-\frac 1{\varepsilon _1(r)}\right]
dr+4\pi \gamma r_s^2  \label{delgmixp4}
\end{equation}
the integral being extended over regions relatively far from the ion. This
would naturally not be the case for an analogous extension of eq. \ref
{delgmix} to consider dielectric saturation, which would result in:

\begin{equation}
\Delta G_{trans}^0=\frac{q^2}2\dint\limits_{r_0}^\infty \frac 1{r^2}\left[
\frac 1{\varepsilon _2(r)}-\frac 1{\varepsilon _1(r)}\right] dr
\label{delgmixp5}
\end{equation}

The value of integral is here according to our experience {\it extremely}
sensitive to assumptions made concerning the structure of the dielectric
media in the neighborhood of the ion.

Returning to our discussion above, it seems that hydration in the organic
phase should only occur for the smaller ions. However, the model given above
is only adequate to describe a complete solvation layer, and hydration by
one or a couple of water molecules may occur. A {\it rough} estimation of
the free energy decrease $\Delta G_{repl}$ that accompanies the replacement
of part of the organic solvation layer by{\it \ a single} water molecule may
be made according to the following argument. We divide the free energy
change related to this hydration process in two parts. One corresponding to
the ion-dipole interactions $\Delta G_{i-dip}$ and another one corresponding
to dipole-dipole interactions $\Delta G_{dip-dip}^0$. The change related to
the ion-dipole interactions can be written as:

\begin{equation}
\Delta G_{ion-dip}^0=-\frac{q^2}{n_{H_2O}^2\ r_{i-H_2O}^2}\mu _{H_2O}<\cos
\Theta _{H_2O}>+\frac{\rho _{DCE}}{\rho _{H_2O}}\frac{q^2}{n_{DCE}^2\
r_{i-DCE}^2}\mu _{DCE}<\cos \Theta _{DCE}>  \label{delgiondip}
\end{equation}
where $n$ labels the refractive index, $\mu $ the dipole moment, $\rho $ the
number density and $<\cos \Theta >$ the average orientation of water ($H_2O$%
) and 1,2-dichloroethane ($DCE$) dipoles respectively. The ion-dipole
distances denoted as $r_{i-H_2O}$ and $r_{i-DCE}$ can be taken to be equal
to the sum of the ionic radius plus the corresponding molecular radius.\ The
average orientation can be estimated using Langevin's equation $L(x)=\coth
(x)-1/x$ with $x=E\mu /kT.$ Thus, the first term on the r.h.s of eq. \ref
{delgiondip} represents the free energy decrease when the ion is solvated by
a water dipole, and the second term represents the corresponding free energy
increase due to desolvation of the corresponding part of the organic
solvent. The values obtained from this equation are shown in Table 3 for the
different ions considered here. As an estimation for the $\Delta
G_{dip-dip}^0$ term we consider the free energy change corresponding to the
dissolution of water in 1,2-dichloroethane at the saturation concentration,
that can in principle be obtained from solubility data of $H_2O$ in $1,2-DCE.
$ Since these solubility values reported in the literature \cite{polacos}
vary between $0.1$ M and 10$^{-3}$M, we estimated $\Delta G_{dip-dip}^0$ to
be in the range $10-30KJ/mol$. Whatever the accurate value, we see that
within the present approximation practically all the ions considered in
Table 3 should prefer solvation with one water molecule to be anhydrous in
the organic solvent. Moreover, since the resulting values for $\Delta
G_{repl}^0=\Delta G_{ion-dip}^0+\Delta G_{dip-dip}^0$ are of the same order
of magnitude that a reasonable calculation or free energy of transfer
without hydration would yield, we expect that even solvation with a single
water molecule should deliver a meaningful contribution to this process.

A more quantitative approach would require a simulation based on realistic
potentials describing the ion-water, ion- 1,2-DCE and all the intermolecular
interactions. In this respect, it is worth mentioning here the interesting
results of the simulation performed by Benjamin\cite{Benjamin92}concerning
the transfer of an ion from a nonpolar to a polar phase. In the case where
the liquid-liquid interface is not forced to be sharp but a wider transition
region is allowed, an important part of the polar solvent solvation shell is
already formed around the ion when it is still on the nonpolar side of the
liquid-liquid interface. It would be thus worthwhile to perform this type of
simulation with more specific potentials for the different ions. This would
provide an important test to the present ideas. Determination of standard
free energies for the transfer of alkaline-earth ions will be also helpful
to distinguish between a WP or a ''dry '' transfer . In fact, any
electrostatic estimation of the standard Gibbs energies of the dry transfer
should scale in a first approximation with $q^2$, as predicted by Born's
equation. On the other hand, replacing eqn. \ref{rsmin} into eqn. \ref
{delgmin} yields:

\begin{equation}
\Delta G_{trans}^{0\prime \text{ }\min }=\frac 32q^{(4/3)}(2\gamma \pi
)^{(1/3)}\left[ \frac 1{\varepsilon _2}-\frac 1{\varepsilon _1}\right]
^{(2/3)}  \label{delgmin2}
\end{equation}

Thus, the WP transfer free energies are expected to scale with $q^{(4/3)}$,
which is much weaker than the $q^2$ dependence given above.

We can summarize the present results saying that we have clarified some
aspects of direct ion transfer across the water/1,2 dichloroethane by using
a very simple model based on thermodynamic considerations. We believe that
ion solvation by water molecules may occur {\it in some particular cases} in
the organic phase, delivering an important contribution to the Gibbs free
energy of ion transfer between the aqueous and the organic phase. This
particular type of transfer should be favored in the case of {\it highly
charged small ions, at interfaces with a relatively low surface tension and
a large difference between the reciprocal of the corresponding dielectric
constants}.

{\bf Acknowledgments}{\Large \ }

{\normalsize \ Financial support from the Secretar\'{\i}a de Ciencia y
Tecnolog\'{\i}a de la Universidad Nacional de C\'{o}rdoba and the Consejo
Nacional de Investigaciones Cient\'{\i}ficas y T\'{e}cnicas, language
assistance from Pompeya Falcon and a fellowship (C.S.) from the Consejo de
Investigaciones de la Provincia de C\'{o}rdoba are gratefully acknowledged. }
\newpage
Table 1. Standard Gibbs transfer energies of univalent ions from water to
1,2-dichloroethane (as quoted in Table 2 of ref. \cite{Sabela}).

\begin{center}
\begin{tabular}{|c|c|}
\hline
Ion & $\Delta G_{tr}^0/kJ\ mol^{-1}$ \\ \hline
H$^{+}$ & $53$ \\ \hline
Li$^{+}$ & $57$ \\ \hline
Na$^{+}$ & $57$ \\ \hline
K$^{+}$ & $50$ \\ \hline
Rb$^{+}$ & $42$ \\ \hline
Cs$^{+}$ & $35$ \\ \hline
Cl$^{-}$ & $51$ \\ \hline
Br$^{-}$ & $43$ \\ \hline
I$^{-}$ & $33$ \\ \hline
\end{tabular}
\end{center}

Table 2. Calculated radii of solvated ion and standard Gibbs energies for
ion transfer surrounded by a layer of water. The values labelled with (Born)
were calculated using Born's solvation model while the values labelled with
(Sat) were calculated taking into account dielectric saturation through eqn. 
\ref{delgmixp3}. The parameters employed in the numerical minimization of eq.%
\ref{delgmixp3} were: $n_{H_2O}=1.3329$, $n_{DCE}=1.4448$ , $\varepsilon
_{H_2O}(bulk)=78.54,$ $\varepsilon _{DCE}(bulk)=10.23$, $\gamma
_{H_2O/DCE}=25.5$ dyn/cm,

\begin{center}
\begin{tabular}{lllll}
Valence / $Z$ & $r_s$ & $r_s$ & $\Delta G_{trans}^{\prime \min }/kJ\
mol^{-1} $ & $\Delta G_{trans}^{\prime \min }/kJ\ mol^{-1}$ \\ 
& $(Born)$ & $(Sat)$ & $(Born)$ & $(Sat)$ \\ 
$1$ & 2.48 & 3.44 & 35.6 & 45.6 \\ 
$2$ & 3.94 & 5.22 & 89.7 & 108.8
\end{tabular}
\end{center}

Table 3. Ion-dipole contribution to the free energy associated with the
replacement of part of the organic solvation layer of an ion by{\it \ a
single} water molecule . All values are in $kJ\ mol^{-1}$.

\begin{center}
\begin{tabular}{|c|c|}
\hline
Ion & $\Delta G_{ion-dip}/kJ\ mol^{-1}$ \\ \hline
Li$^{+}$ & -117.4 \\ \hline
Na$^{+}$ & -91.1 \\ \hline
K$^{+}$ & -69.0 \\ \hline
Rb$^{+}$ & -62.4 \\ \hline
Cs$^{+}$ & -54.5 \\ \hline
F$^{-}$ & -69.0 \\ \hline
Cl$^{-}$ & -49.9 \\ \hline
Br$^{-}$ & -45.5 \\ \hline
I$^{-}$ & -39.5 \\ \hline
\end{tabular}
\end{center}

\bigskip\ 

\bigskip\ 

\begin{center}
{\normalsize {\Large {\bf Figure captions} }}
\end{center}

{\normalsize \newcounter{fig} \TeXButton{list}
{\begin{list}{\bf Figure \arabic{fig}:}{\usecounter{fig}\sl}
\item Model for the direct transfer of an alkaline ion of radius $r_0$ through the water/organic interface. We allow for the possibility that the ion carries with it a solvation sheet which is essentially similar to that existing in the aqueous phase. The conditions under which this may occur are discussed in the text.
\item Qualitative plot of the transfer free energy of an ion with an aqueous solvation sheet ($\Delta G_{trans}^{\prime })$ , as a function of the radius $r_s$ of a sphere containing the ion+aqueous layer. a) The horizontal line labelled ($\Delta G_{trans}^{})$ indicates the transfer free energy of an ion without the aqueous layer. These schemes correspond to the conditions: a)$\frac{\gamma r_0^3}{q^2}<\frac 1{54\pi }\left[ \frac 1{\varepsilon _2}-\frac 1{\varepsilon _1}\right] $; b)$\frac{\gamma r_0^3}{q^2}=\frac 1{54\pi }\left[ \frac 1{\varepsilon _2}-\frac 1{\varepsilon _1}\right] $; c)$\frac{\gamma r_0^3}{q^2}>\frac 1{54\pi }\left[ \frac 1{\varepsilon _2}-\frac 1{\varepsilon _1}\right] $
\item Electrochemical standard free energies for the transfer of univalent ions from water to 1,2-dichloroethane. Empty circles indicate cations, while filled circles denote anions. The reciprocal radii were calculated from a)ionic radii as reported in \cite{Bockris} b) corrected radii as reported by Rashin and Honig \cite{Rashin&Honig85}.
\end{list}
}}

{\normalsize \ }

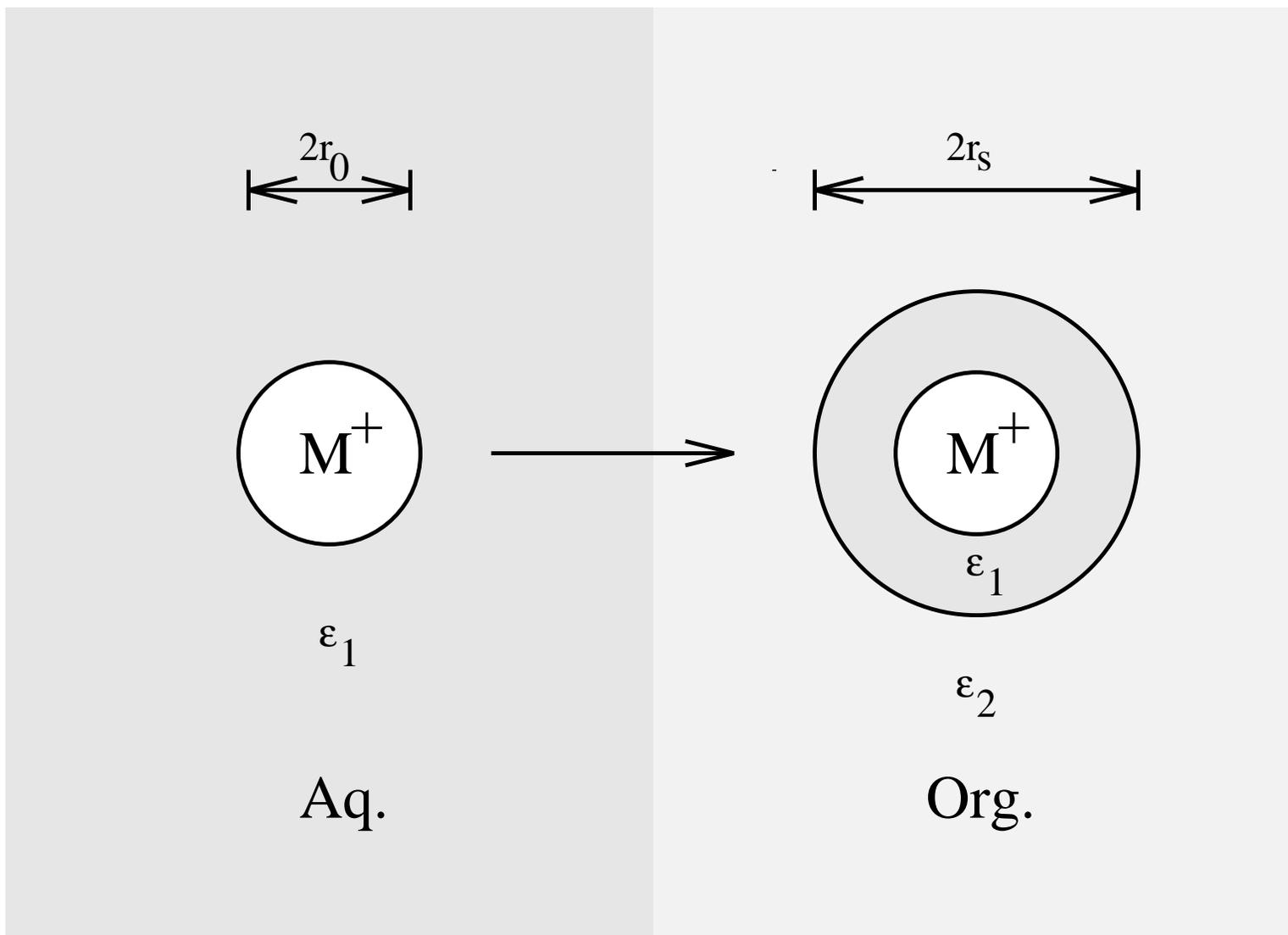
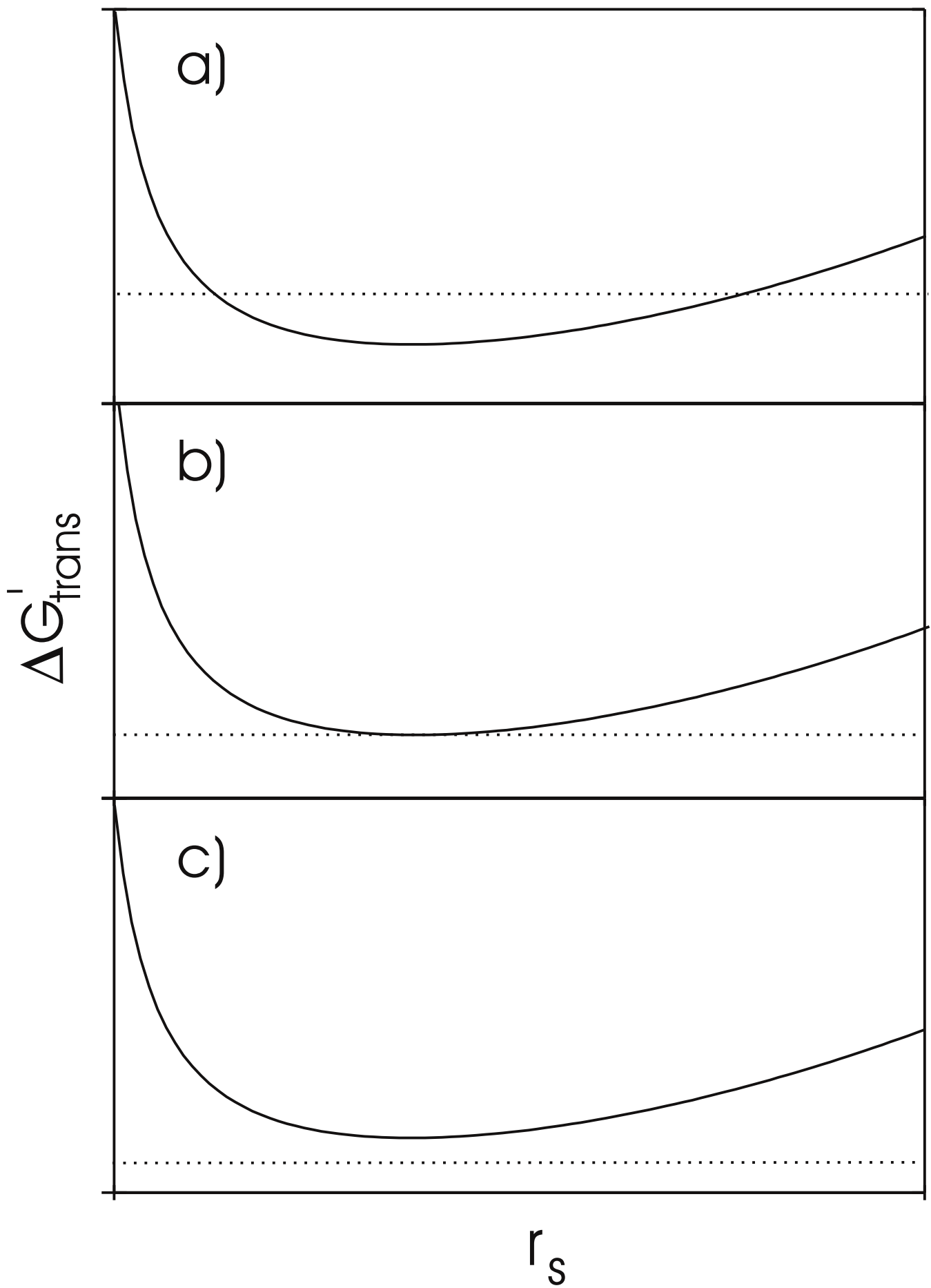
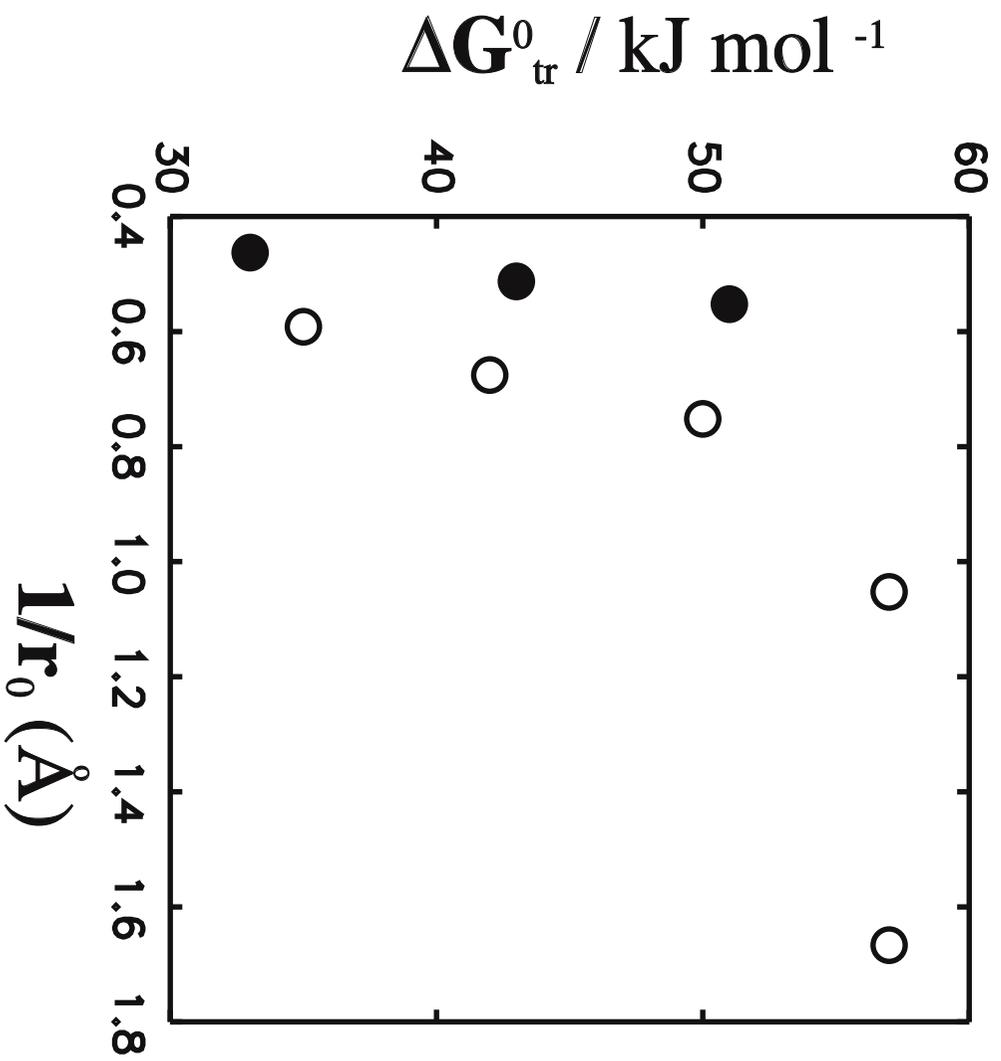
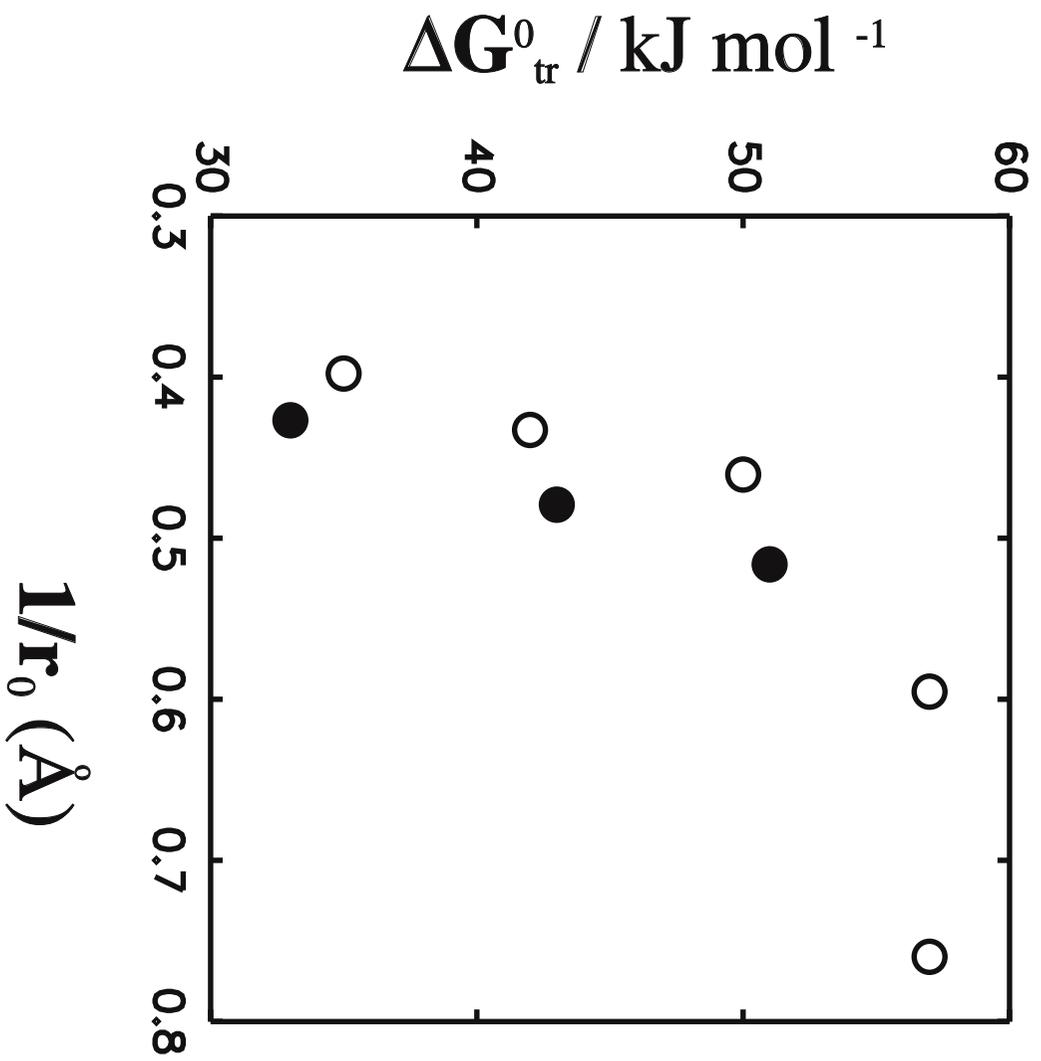
\end{document}